\documentclass[acmtog, nonacm = true]{acmart}
\renewcommand\footnotetextcopyrightpermission[0]{} 
\acmSubmissionID{papers 790s1}

\AtBeginDocument{%
  \providecommand\BibTeX{{%
    \normalfont B\kern-0.5em{\scshape i\kern-0.25em b}\kern-0.8em\TeX}}}

\usepackage{amsmath, amsthm, mathtools}
\usepackage{breqn}
\usepackage{subcaption}
\usepackage{cleveref}
\usepackage{enumitem}
\usepackage{makecell}

\begin{document}

\title{Differentiable Neural Radiosity}

\author{Saeed Hadadan}
\orcid{1234-5678-9012-3456} 
\affiliation{%
  \institution{University of Maryland, College Park}
  \city{College Park}
  \state{MD}
  \postcode{20740}
  \country{USA}}
\email{saeedhd@umd.edu}

\author{Matthias Zwicker}
\affiliation{%
 \institution{University of Maryland, College Park}
 \city{College Park}
 \state{MD}
 \country{USA}}
\email{zwicker@cs.umd.edu}

\renewcommand\shortauthors{Hadadan, S. et al}

\begin{abstract}
  We introduce Differentiable Neural Radiosity, a novel method of representing the solution of the \emph{differential rendering equation} using a neural network. Inspired by neural radiosity techniques, we minimize the norm of the residual of the differential rendering equation to directly optimize our network. The network is capable of outputting continuous, view-independent gradients of the radiance field with respect to scene parameters, taking into account differential global illumination effects while keeping memory and time complexity constant in path length. To solve inverse rendering problems, we use a pre-trained instance of our network that represents the differential radiance field with respect to a limited number of scene parameters. In our experiments, we leverage this to achieve faster and more accurate convergence compared to other techniques such as Automatic Differentiation, Radiative Backpropagation, and Path Replay Backpropagation. 
\end{abstract}


\begin{teaserfigure}
  \includegraphics[width=\textwidth]{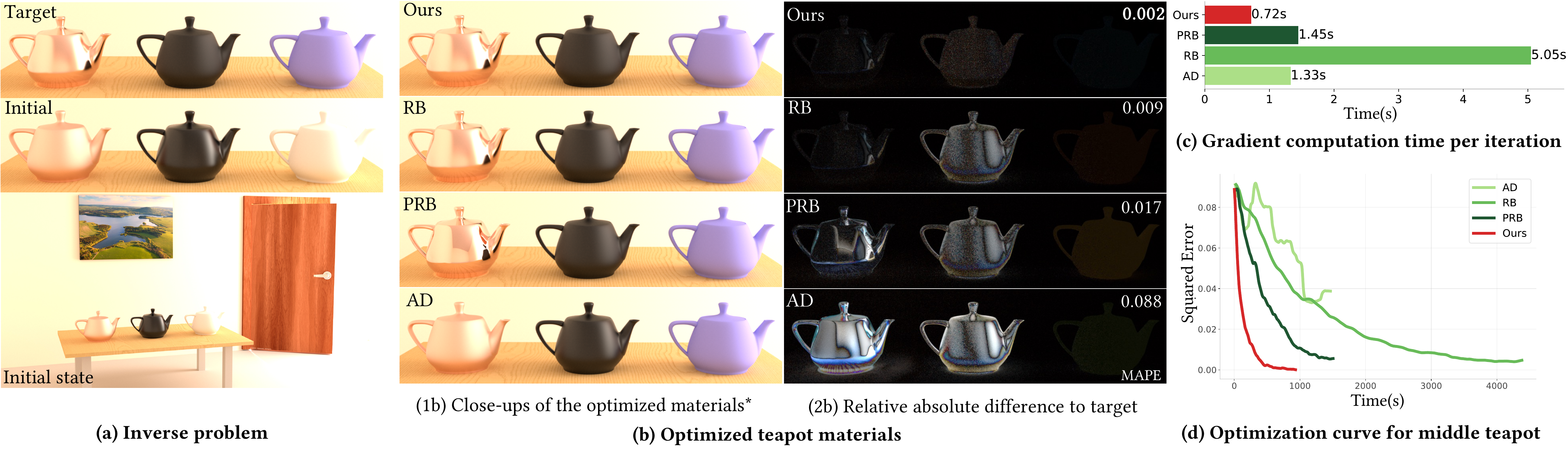}
  \caption{Inverse rendering using Differentiable Neural Radiosity, compared to Automatic Differentiation, Radiative Backpropagation, and Path Replay Backpropagation.  (a) We optimize one BRDF parameter per teapot, roughness for the left and middle teapots and diffuse albedo for the right teapot. 
  (b) Renderings of the optimized materials using our method are  more faithful to the target state than the previous state-of-the-art. (c) Given a pre-trained network, gradient computation time during inverse optimization experiences considerable speed-ups using our method. (d) The squared error of the BRDF parameters (here roughness parameter of the middle teapot) diminishes faster than other methods against time (details in Figure ~\ref{fig:optim_curves}). *Close-ups are only for visualization purpose; the target view is the image (a) bottom, using a more distant sensor.}
  \label{fig:teaser}
\end{teaserfigure}

\maketitle

\section{Introduction}

Differentiable rendering is the problem of computing derivatives of a rendering process with respect to scene parameters such as BRDFs, geometry, illumination, or camera parameters. Differentiable rendering is attractive because it enables gradient-based optimization in inverse rendering problems, where scene parameters are optimized such that the rendering process produces a target image (or multiple target images from different viewpoints). In combination with neural network-based 3D representations, differential rendering is popular because it enables end-to-end training of 3D neural representations with only images as training data. 

Here we are focusing on differentiable rendering for algorithms that solve the rendering equation, with the ultimate goal to eventually enable inverse rendering based on real world photographs under large numbers of unknown scene parameters. Previous techniques build on Monte Carlo path tracing to solve the rendering equation, and differentiable rendering can be achieved using generic Automatic Differentiation (AD) techniques, or with specialized algorithms such as Radiative Backpropagation (RB) or Path Replay Backpropagation (PRB). However, these techniques can be computationally costly and they often cannot provide highly accurate results when computation time is limited as shown in Figure~\ref{fig:teaser}.

In this paper, we propose a novel technique that is inspired by the Neural Radiosity approach by Hadadan et al.~\shortcite{hadadan2021neural} to solve the rendering equation. Our key idea is to represent the entire differential radiance field, that is, the derivative of the radiance field with respect to a set of scene parameters, using a neural network. We then train this network to satisfy the differential rendering equation introduced by Nimier-David et al.~\shortcite{radiative}, and our training scheme is similar in spirit as in Neural Radiosity. We show that given a trained network for an initial scene, we can solve inverse rendering problems to fit BRDF parameters more efficiently and more accurately than previous work. In addition, our approach is the first that produces a continuous view-independent differential radiance field for given scene parameters, instead of only sampling derivatives for a discrete set of rays.

\section{Related Work}

\paragraph{Neural Network Techniques for Realistic Rendering.}

Neural networks can be leveraged in various ways for realistic rendering. Vogels et al.~\shortcite{Vogels18DKP} used neural networks for post-processing of rendered images to achieve denoising, anti-aliasing and super-resolution. Neural networks have also been utilized at the core of realistic rendering algorithms. For example, M\"uller et al.~\shortcite{Muller2019NIS} and Zhen and Zwicker~\shortcite{Zhen2019LIPSS} use neural networks to learn to importance sample the radiance distribution during Monte Carlo rendering. 
In addition, M\"uller et al.~\shortcite{NCV} train neural networks to learn control variates, which is a variance reduction technique for solving the rendering equation~\cite{kajiya_rendering_equation}. Neural networks can also directly represent the solution of the rendering equation as proposed in two concurrent works, Neural Radiosity~\cite{hadadan2021neural} and Neural Radiance Caching~\cite{Mueller2021NRC}. In these methods, the benefit of storing the radiance function in a network is that the network can be updated with limited computational effort under incremental changes in the rendered scenes.

This paper aims at leveraging neural networks to learn view-independent, continuous differential radiance fields, which is to the best of our knowledge unprecedented. We use a single neural network to represent the \emph{differential radiance function} w.r.t arbitrary scene parameters. In a similar approach to Neural Radiosity, we optimize our network parameters directly by minimizing the norm of the residual of the \emph{differential} rendering equation. We benefit from the re-renderability of the radiance network as in Neural Radiosity. 

\paragraph{Differentiable Rendering with Indirect Effects.}
Inverse rendering problems in computer vision and graphics heavily rely on differentiable rendering to reconstruct a set of scene parameters (geometry, reflectance properties, camera positions, etc.) from images. Most computer vision techniques rely on differentiable rasterization~\cite{versatile_scene_model,OpenDR, Kato_2018_CVPR, liu2019soft, petersen2019pix2vex, laine2020modular}. These methods simply ignore indirect illumination effects, which require a full solution of the rendering equation. 
Techniques to differentiate the rendering equation while accounting for indirect effects have been proposed in prior work. 
Specifically, \emph{Radiative Backpropagation} (RB)~\cite{radiative} differentiates the rendering equation to obtain a \emph{differential rendering equation}, which describes scattering and emission of \emph{differential light}. The equation reveals that \emph{differential light} travels through a scene similarly as regular light does. 
RB proposes an adjoint approach for differentiable rendering, which is more efficient than naive Automatic Differentiation of a Monte Carlo path tracer. 
RB's main shortcoming is that its time complexity is quadratic in path length, since at every scattering event during the adjoint phase, it requires to estimate incident radiance by building another complete light path. \emph{Path Replay Backpropagation} (PRB) \cite{path_replay} improves on RB and has linear complexity in path length. It is based on a two-pass approach, by first performing primal path tracing and storing auxiliary information, which can then be used to invert the process during the second phase. 
Our approach solves the issue of quadratic time in RB by leveraging neural networks to serve as a form of cached radiance and differential radiance.

\paragraph{Visibility-related Discontinuities.}

Differentiating the rendering equation without taking into account discontinuities can lead to incorrect gradients, 
and discontinuities due to occlusions 
need special treatment. Li et al.~\shortcite{edgesampling} tackle this issue by separating out the rendering equation into continuous and discontinuous parts and propose a silhouette 
edge sampling approach. Loubet et al.~\shortcite{reparam} propose a reparametrization of the rendering integrals such that
the positions of discontinuities do not depend on the scene parameters. For simplicity, in this paper we ignore  discontinuities and focus on solving the differential rendering equation as a first step. 


\section{Background}

\subsection{(Differential) Rendering Equation}

Realistic rendering algorithms compute a set of measurements $I_k$ where $k$ corresponds to a pixel, given by the measurement equation
\begin{dmath}
\label{eq:measurement-equation}
I_k = \int_{\mathcal{A}}\int_{\mathcal{H}^2} W_k(x,\omega) L(x,\omega) dx d\omega_i^{\perp},
\end{dmath}
where $L$ is the incident radiance at location $x$ and direction $\omega$ on the pixel, and $W_k$ is the importance of pixel $k$. As radiance $L$ remains constant along unoccluded rays, incident radiance at a pixel location and direction is equal to the outgoing radiance from the nearest surface along the ray. The outgoing radiance at surfaces can be computed using the rendering equation \cite{kajiya_rendering_equation},
\begin{dmath}
\label{eq:rendering-equation}
L(x,\omega_o) = E(x,\omega_o) + \int_{\mathcal{H}^2} f(x,\omega_i, \omega_o) L(x'(x,\omega_i),-\omega_i), d\omega_i^{\perp},
\end{dmath}
which describes energy conservation in scattering events, i.e. the outgoing radiance at every location $x$ and direction $\omega_o$ can be described as emission of the surface $E(x, \omega_o)$ plus scattering of incident radiance. The integral on the right hand side describes light scattering at incidents directions $\omega_i$ over the projected solid angle hemisphere $\mathcal{H}^2$. Function $f$ here represents bidirectional reflectance distribution function (BSDFs).

\begin{figure}[t]
  \centering
\subcaptionbox{Primal rendering}{  
\includegraphics[width = 0.47\columnwidth]{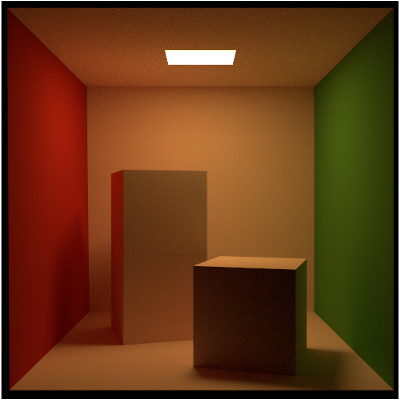}
}
\subcaptionbox{RB}{  
\includegraphics[width = 0.47\columnwidth]{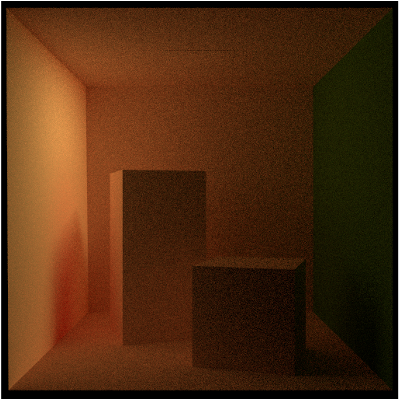}
}

\subcaptionbox{LHS}{  
\includegraphics[width = 0.47\columnwidth]{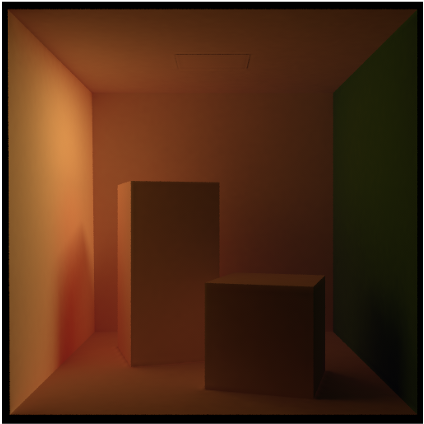}
}
\subcaptionbox{RHS}{  
\includegraphics[width = 0.47\columnwidth]{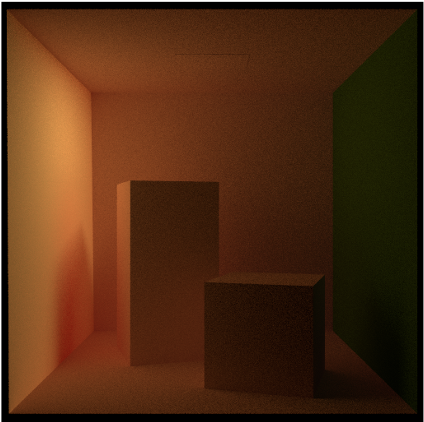}
}
  \caption{Primal and differential renderings w.r.t the red wall diffuse albedo for \emph{cbox}. Similar to primal radiance,  differential radiance scatters through the scene and causes differential global illumination indicating how slight perturbations in the red color affects other parts of the scene. LHS/RHS renderings of our solution are also included to show their agreement with Radiative Backpropagation.}
  \label{fig:cbox}
\end{figure}

Nimier-David et al.~\shortcite{radiative} differentiate the above equations w.r.t to an arbitrary set of scene parameters $p = (p_1, ... , p_n)$. For simplicity,  we use $\partial_p$ to represent $\partial / \partial_p$. Note that variables preceded by $\partial_p$ imply a vectorized gradient w.r.t each parameter. By assuming a static camera where $\partial_{p}W_k = 0$, we can differentiate Equation \ref{eq:measurement-equation} as,
\begin{dmath}
\label{eq:differential-measurement-equation}
\partial_{p}I_k = \int_{\mathcal{A}}\int_{\mathcal{H}^2} W_k(x,\omega) \partial_p L(x,\omega) dx d\omega_i^{\perp},
\end{dmath}
which describes the relationship between \emph{differential measurement} $\partial_{p}I_k$ and \emph{differential radiance} $\partial_{p}L$. Differential radiance $\partial_{p}L$ in turn can be found by differentiating Equation \ref{eq:rendering-equation}, 
\begin{dmath}
\label{eq:differntial-rendering-equation}
\partial_{p}L(x,\omega_o) = \partial_{p}E(x,\omega_o) + \int_{\mathcal{H}^2} f(x,\omega_i, \omega_o) \partial_{p}L(x'(x,\omega_i),-\omega_i) d\omega_i^{\perp}+
\int_{\mathcal{H}^2} \partial_{p}f(x,\omega_i, \omega_o) L(x'(x,\omega_i),-\omega_i) d\omega_i^{\perp},
\end{dmath}
which we call the \emph{differential rendering equation}. This equation explains the scattering of differential radiance in a similar manner to regular radiance (Figure ~\ref{fig:cbox}). More specifically, the first term describes how differential radiance is \emph{emitted} from surfaces whose emission is dependent on the scene parameters $p$. The second term means that differential radiance \emph{scatters} on surfaces based on their BRDFs, just like regular radiance in the rendering equation. The new third term represents additional differential \emph{emission} from the surface if its BRDF function changes with perturbations of scene parameters $p$. This term is dependent on the incident radiance $L$ at $(x,\omega_i)$, which implies computing $\partial_p L$ depends on computing $L$ also.

\subsection{Neural Radiosity}

Neural Radiosity \cite{hadadan2021neural} is an algorithm to find a solution of the rendering equation (Equation \ref{eq:rendering-equation}) using a single neural network. More formally, the radiance function $L(x,\omega_o)$ in Equation \ref{eq:rendering-equation} is represented by a neural network with a set of parameters $\phi$, as $L_{\phi}(x,\omega_o)$. The parameters $\phi$ of this network can be directly optimized in a self-training approach by minimizing the norm of the \emph{residual} of the rendering equation. The residual $r_{\phi}(x,\omega_o)$ is defined as, 
\begin{dmath}
\label{eq:renderingresidual}
    r_{\phi}(x,\omega_o) =  L_{\phi}(x,\omega_o) - E(x,\omega_o) \nonumber \\
     - \int_{\mathcal{H}^2} f(x,\omega_i, \omega_o) L_{\phi}(x'(x,\omega_i),-\omega_i) d\omega_i^{\perp},
\end{dmath}
which is simply the difference of the left and right-hand sides of Equation \ref{eq:rendering-equation} when the radiance function $L$ is substituted by $L_{\phi}$. This neural network takes a location $x$ and outgoing direction $\omega_o$ as input (along with extra encoding information, see \cite{hadadan2021neural}) and returns the outgoing radiance. Such a pre-trained network serves as a compact, re-renderable, and view-independent solution of the rendering equation. 

\section{Solving the Differential Rendering Equation}

Similar to Neural Radiosity, we propose to use neural network-based solvers to find the solution of the \emph{differential rendering equation}. We call this Differentiable Neural Radiosity. Let us denote a differential radiance distribution  $\partial_{p}L_{\theta}(x,\omega_o)$ as the unknown in Equation \ref{eq:differntial-rendering-equation}, given by a set of network parameters $\theta$. Additionally, we define a \emph{residual} $r_{\theta}$ as the difference of the left and right hand side of Equation \ref{eq:differntial-rendering-equation},
\begin{dmath}
\label{eq:differntial-residual}
r_{\theta}(x,\omega_o) = \partial_{p}L_{\theta}(x,\omega_o) - \partial_{p}E(x,\omega_o) - \int_{\mathcal{H}^2} f(x,\omega_i, \omega_o) \partial_{p}L_{\theta}(x'(x,\omega_i),-\omega_i) d\omega_i^{\perp}-
\int_{\mathcal{H}^2} \partial_{p}f(x,\omega_i, \omega_o) L_{\phi}(x'(x,\omega_i),-\omega_i) d\omega_i^{\perp},
\end{dmath}
where $r_{\theta}$ depends on the parameters $\theta$ of the differential radiance function $\partial_{p}L_{\theta}$. Also, the primal radiance can be represented by a constant parameter set $\phi$ in $L_{\phi}$ which is independent of $\theta$.      

We define our loss as the L2 norm of the residual,
\begin{align}
\label{eq:differential-renderingloss}
\mathcal{L}(\theta) &= \left\| r_{\theta}(x,\omega_o) \right\|_2^2 \nonumber \\
&= \int_{\mathcal{M}}\int_{\mathcal{H}^2} r_{\theta}(x,\omega_o)^2 dxd\omega_o, 
\end{align}
where $\mathcal{M}$ means integration over all scene surfaces. We propose to minimize $\mathcal{L}(\theta)$ using stochastic gradient descent.

\paragraph{Monte Carlo Estimation.} The Monte Carlo estimation of the residual norm is  
\begin{align}
    \mathcal{L}(\theta) \approx \frac{1}{N}\sum_{j=1}^{N} \frac{r_{\theta}(x_j,\omega_{o,j})^2}{p(x_j,\omega_{o,j})},
    \label{eq:lossMC}
\end{align}
where $N$ is the number of samples, $x_j$ and $\omega_{o,j}$ are the surface location and the outgoing direction samples, taken from a distribution with density $p(x,\omega)$.

The Monte Carlo estimation of the incident integral for any $r_{\theta}(x_j,\omega_{o,j})$ is
\begin{align}
\label{eq:incidentintegral-MCestimation}
& r_{\theta}(x_j,\omega_{o,j}) = \partial_{p}L_{\theta}(x_j,\omega_{o,j})-\partial_{p}E(x_j,\omega_{o,j}) \nonumber \\
&-\frac{1}{M} \sum_{k=1}^{M} \frac{f(x_j,\omega_{i,j,k}, \omega_{o,j})  \partial_{p}L_{\theta}(x'(x_j,\omega_{i,j,k}),-\omega_{i,j,k})}{p(\omega_{i,j,k})} \nonumber \\
&-\frac{1}{Z} \sum_{l=1}^{Z} \frac{\partial_{p}f(x_j,\omega_{i,j,l}, \omega_{o,j}) L_{\phi}(x'(x_j,\omega_{i,j,l}),-\omega_{i,j,l})}{p(\omega_{i,j,l})}.
\end{align}
The notation $\omega_{i,j,k}$ and $\omega_{i,j,l}$ indicates that each sample $x_j,\omega_{o,j}$ has its own set of samples of $M$ and $Z$ incident directions $\omega_{i,j,k}$ and $\omega_{i,j,l}$ ($i$ stands for ``incident'', it is not an index).

\paragraph{Gradients}

The Monte Carlo approximation of the gradient of the loss is 
\begin{align}
\label{eq:residualsampling}
\nabla_{\theta} \mathcal{L}(\theta) \approx& \frac{1}{N}\sum_{j=1}^{N} \frac{2 r_{\theta}(x_j,\omega_{o,j}) \nabla_{\theta} r_{\theta}(x_j,\omega_{o,j})}{p(x_j,\omega_{o,j})},
\end{align}
where the gradient of the residual is
\begin{align}
\label{eq:differntial-residual-gradient}
\nabla_{\theta} r_{\theta}(x,\omega_o) = &\nabla_{\theta}(\partial_{p}L_{\theta}(x,\omega_o)) \\
&- \int_{\mathcal{H}^2} f(x,\omega_i, \omega_o) \nabla_{\theta}(\partial_{p}L_{\theta}(x'(x,\omega_i),-\omega_i)) d\omega_i^{\perp}. \nonumber
\end{align}
This gradient will be approximated by the Monte Carlo estimation of the incident integral,
\begin{align}
\label{eq:graidentincidentintegral-MCestimation}
&\nabla_{\theta} r_{\theta}(x_j,\omega_{o,j}) = \nabla_{\theta} (\partial_{p}L_{\theta}(x_j,\omega_{o,j})) \nonumber \\
&-\frac{1}{M} \sum_{k=1}^{M} \frac{f(x_j,\omega_{i,j,k}, \omega_{o,j}) \nabla_{\theta} (\partial_{p}L_{\theta}(x'(x_j,\omega_{i,j,k}),-\omega_{i,j,k}))}{p(\omega_{i,j,k})}.
\end{align}
All the above gradients can be computed using a deep learning automatic differentiation framework during the optimization process, including the gradient terms $\partial_p f$ and $\partial_p E$.

\section{Inverse Rendering Using Our Method}
\label{inverse-redering}
For inverse rendering, the goal is to optimize a set of scene parameters $p$ using an objective function $z(.)$, which denotes the distance between a candidate image to the reference, and a rendering function $g(.)$. To minimize $z(g(p))$, we need the gradient $\frac{\partial z}{\partial p}$,
\begin{align}
\label{eq:inverse-chain-rule}
\frac{\partial z}{\partial p} = \frac{\partial z}{\partial y}. \frac{\partial y}{\partial p},
\end{align}
where $y$ is a rendered image $y = g(p)$. The term $\frac{\partial z}{\partial y}$ can be interpreted as the gradient of the loss w.r.t pixel values of the candidate image. In most cases, computing this gradient is easy either manually (e.g. if it is $L2$ or $L1$) or using AD (e.g. if it is a neural network). The more challenging part is $\frac{\partial y}{\partial p}$, which is equivalent to the differential measurement vector $[\partial_p I_{0} ... \partial_p I_{n}]$, since we need to differentiate the rendering algorithm. Recall from Equation \ref{eq:differential-measurement-equation} that $\partial_p I_{k}$ is the result of integrating the incident \emph{differential radiance} ${\partial_p L}$ over locations $x$ and directions $\omega$ on the hemisphere at pixel $k$. Therefore, the task breaks down to finding $\partial_p L(x,w)$. In fact, our proposal is to query our neural network ${\partial_p L_{\theta}(x,w)}$ for the differential radiance, as it can serve as a cached representation of the whole differential radiance distribtuion.

With the use of our network, inverse rendering breaks down into the following steps: 
\begin{enumerate}
    \item Compute a non-differentiable candidate primal rendering and its distance to the reference ($L2, L1$, etc.).
    \item Find the derivatives of the loss w.r.t the pixels of the primal image (to get $\frac{\partial z}{\partial y}$).
    \item Compute $\frac{\partial y}{\partial p}$, which is equivalent to the measurement vector $[\partial_p I_{0} ... \partial_p I_{n}]$. To do so, we trace rays from the sensor to find the first hit point and at that point, query our differential radiance network $\partial_p L_{\theta}$. More formally, 
    \begin{equation}
    \label{eq:our-differential-measurement-equation}
    \partial_{p}I_{k,\theta} = \int_{\mathcal{A}}\int_{\mathcal{H}^2} W_k(x,\omega) \partial_p L_{\theta}(x'(x,\omega)) dx d\omega_i^{\perp}.
    \end{equation}
    \item Compute the gradient of the loss w.r.t the parameters $\frac{\partial z}{\partial p}$ by multiplying the gradients from step (2) and (3) as in Equation \ref{eq:inverse-chain-rule}.
    \item Update the scene parameters using the computed gradient using a learning rate and an optimizer (e.g. Adam~\cite{kingma2017adam}).
    \item Fine-tune our networks $\partial_p L_{\theta}$ and $L_{\phi}$ after updating the scene parameters. 
\end{enumerate}

Figure~\ref{fig:pipeline_inverse} summarizes the steps of our pipeline.

\newcommand\teapotw{0.11}

\begin{figure}[t]
    \centering

    \vspace{5pt}

     \makebox[5pt]{\rotatebox{90}{\hspace{15pt} LHS}}
    \includegraphics[width=\teapotw\textwidth]{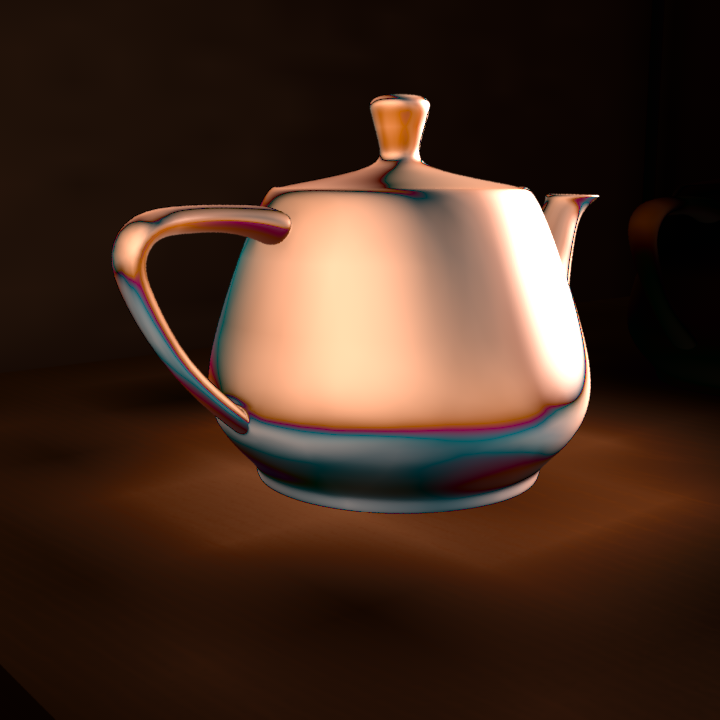}
     \includegraphics[width=\teapotw\textwidth]{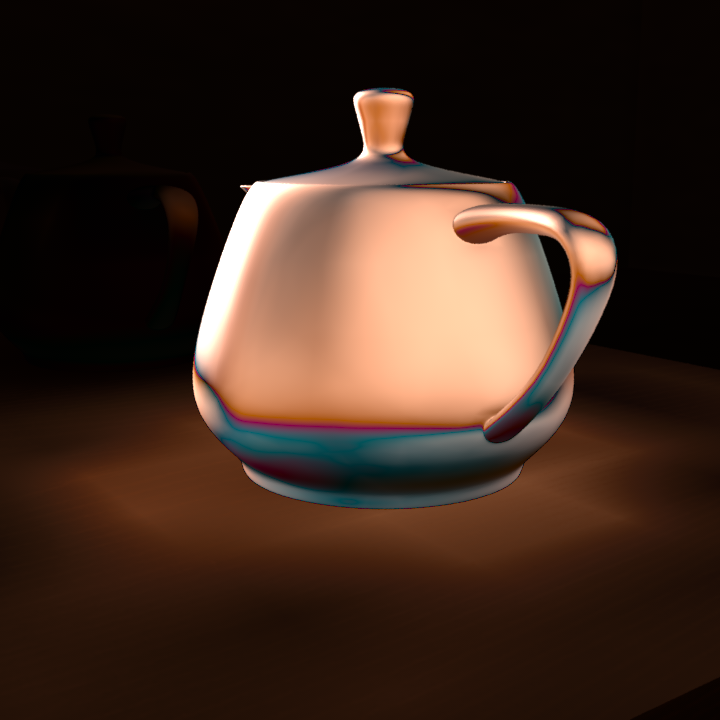}
     \includegraphics[width=\teapotw\textwidth]{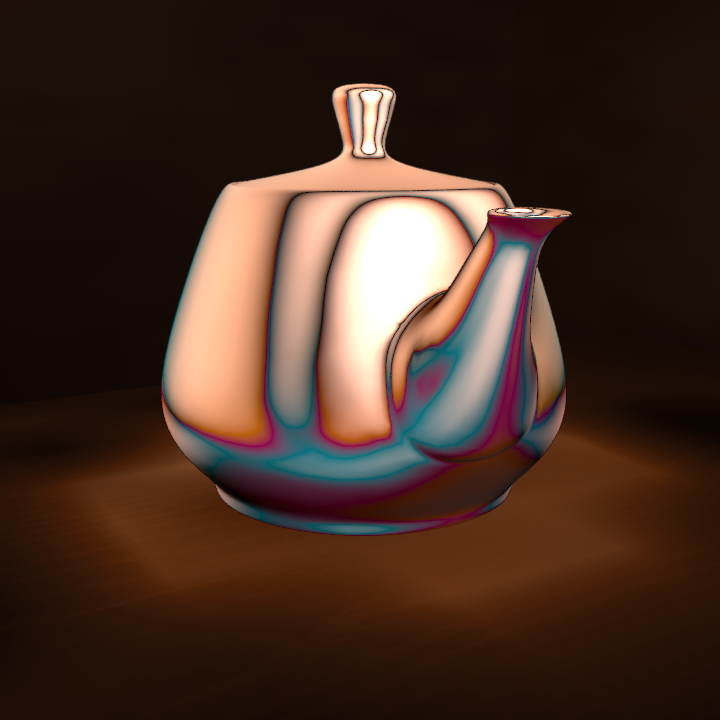}
     \includegraphics[width=\teapotw\textwidth]{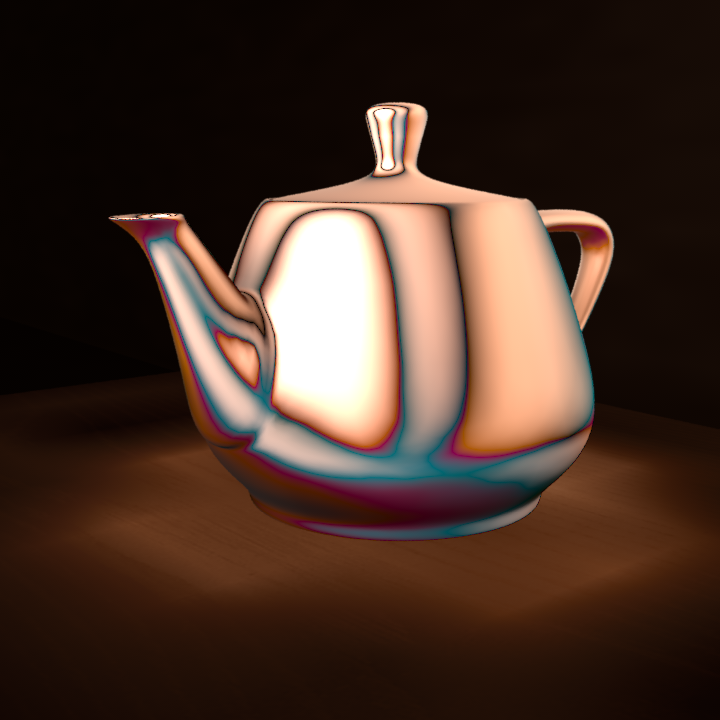}

     \makebox[5pt]{\rotatebox{90}{\hspace{15pt} RHS}}
    \includegraphics[width=\teapotw\textwidth]{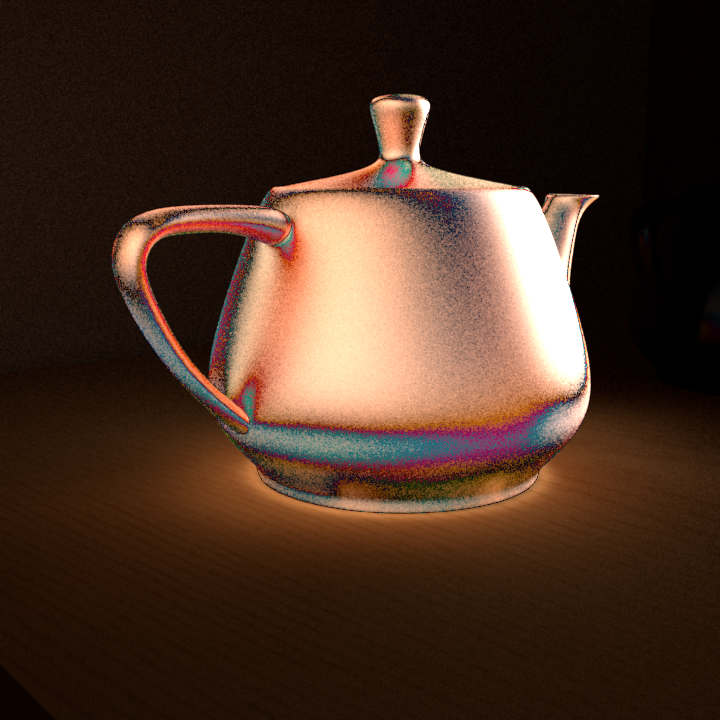}
     \includegraphics[width=\teapotw\textwidth]{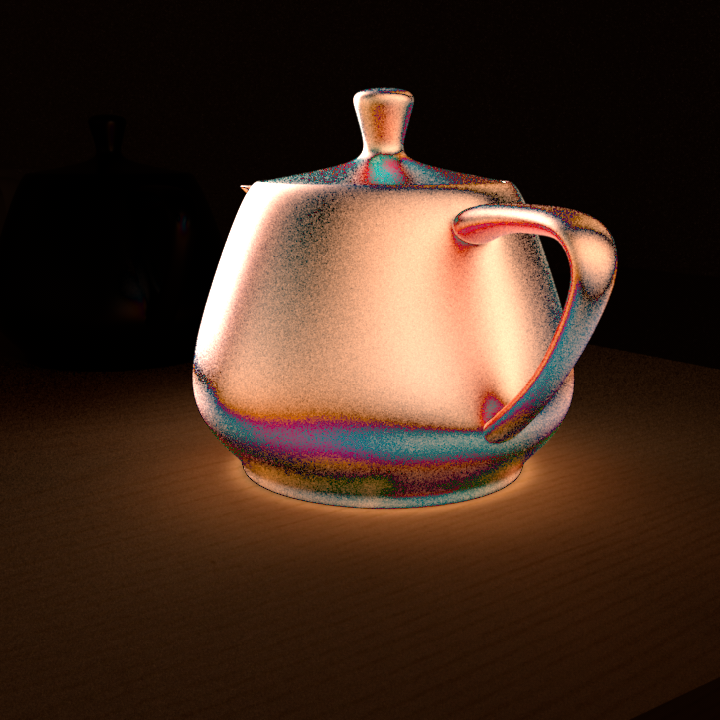}
     \includegraphics[width=\teapotw\textwidth]{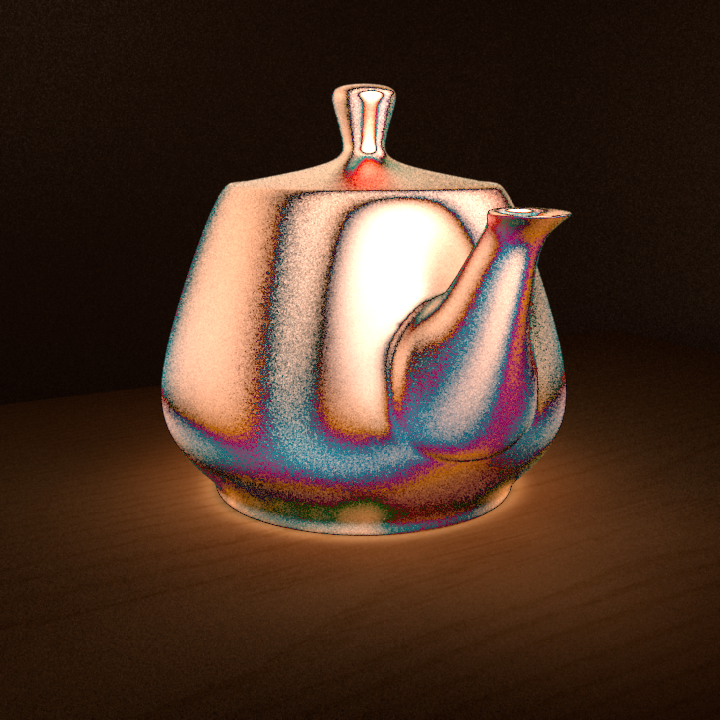}
     \includegraphics[width=\teapotw\textwidth]{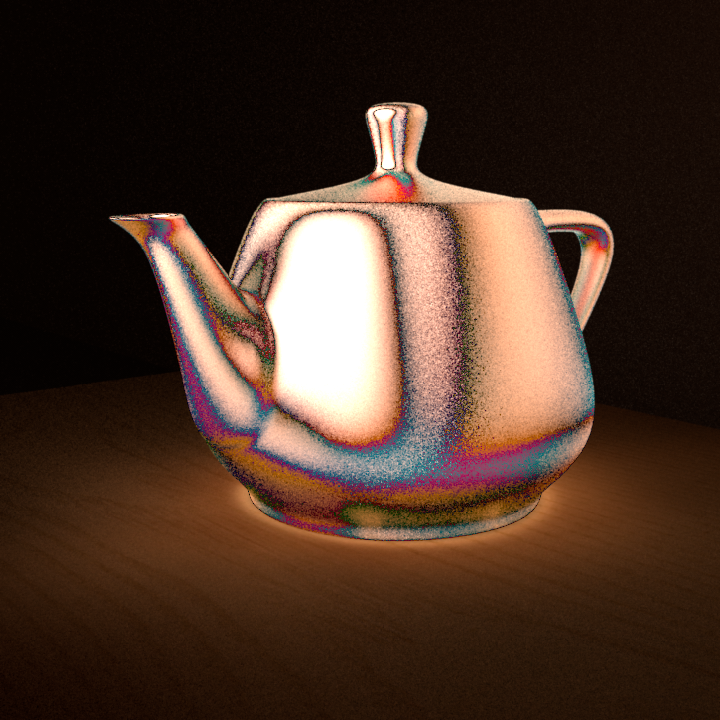}

     \makebox[5pt]{\rotatebox{90}{\hspace{15pt} RB}}
    \includegraphics[width=\teapotw\textwidth]{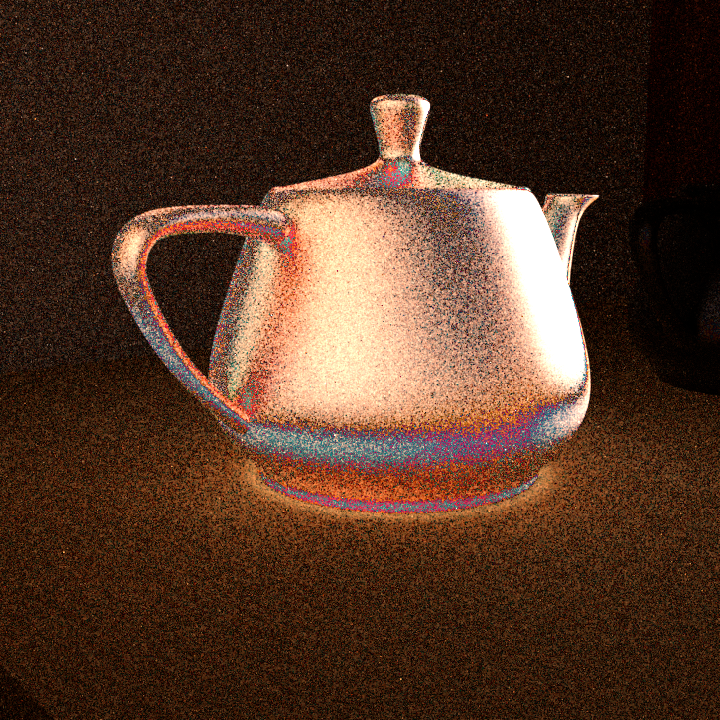}
     \includegraphics[width=\teapotw\textwidth]{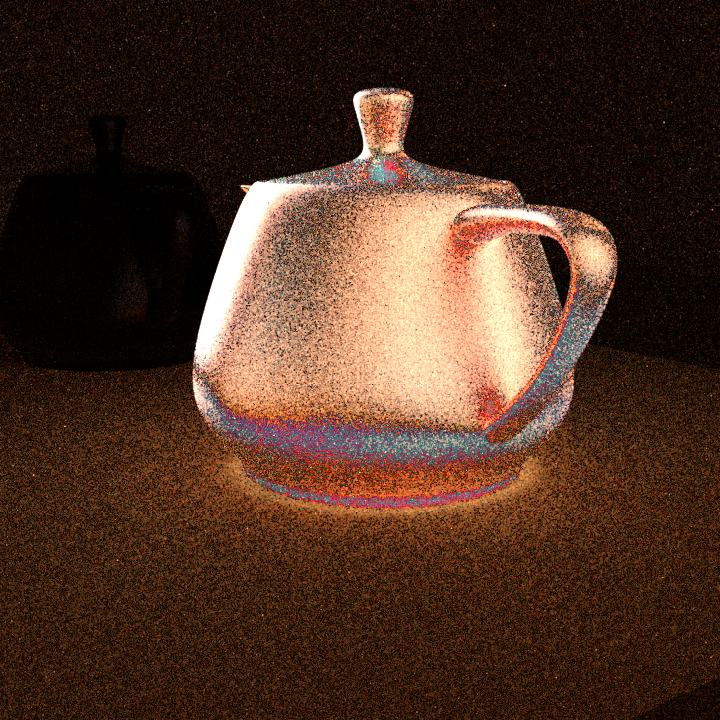}
     \includegraphics[width=\teapotw\textwidth]{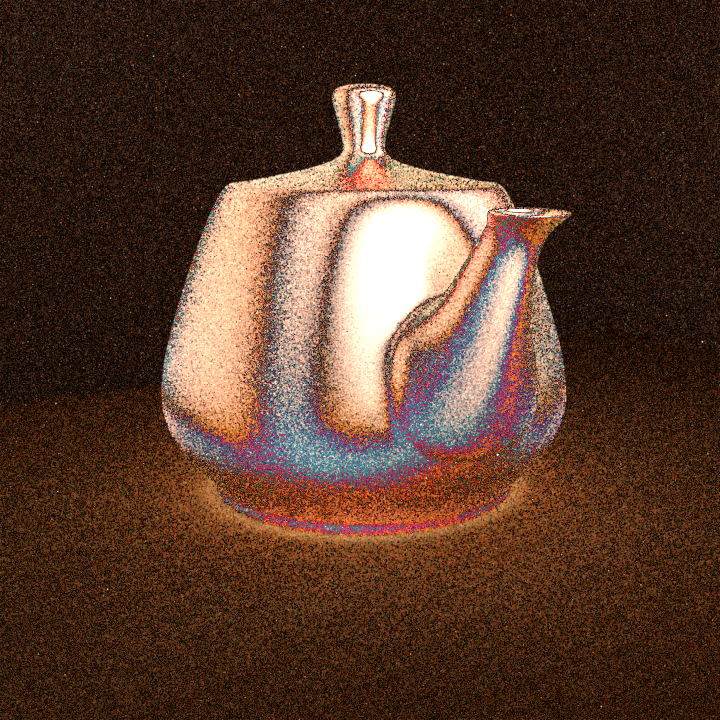}
     \includegraphics[width=\teapotw\textwidth]{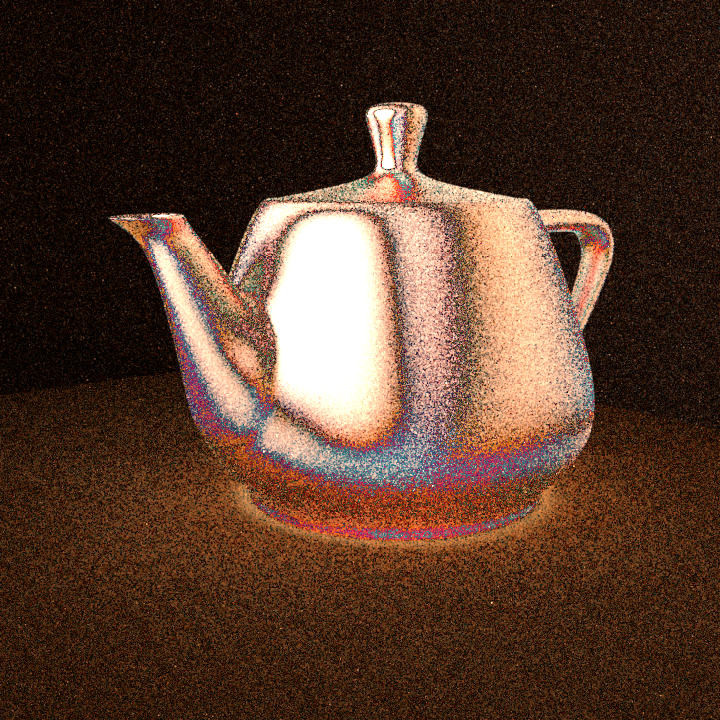}

    \caption{Multi-view renderings of our view-independent solution of the differential rendering equation compared to RB. We show the derivative with respect to the teapot's roughness (the left teapot in Figure ~\ref{fig:teaser}a). Here, roughness = 0.4 as in the initial state in Figure ~\ref{fig:teaser}a. Using 8 spp for LHS and 2048 for RHS and RB.
    \label{fig:multiview}
    }
\end{figure}

\section{Implementation}


\subsection{Training}

\begin{figure*}[ht]
    \subcaptionbox{Computing our loss function during training \label{fig:pipeline_training}}{  
    \includegraphics[width = \textwidth]{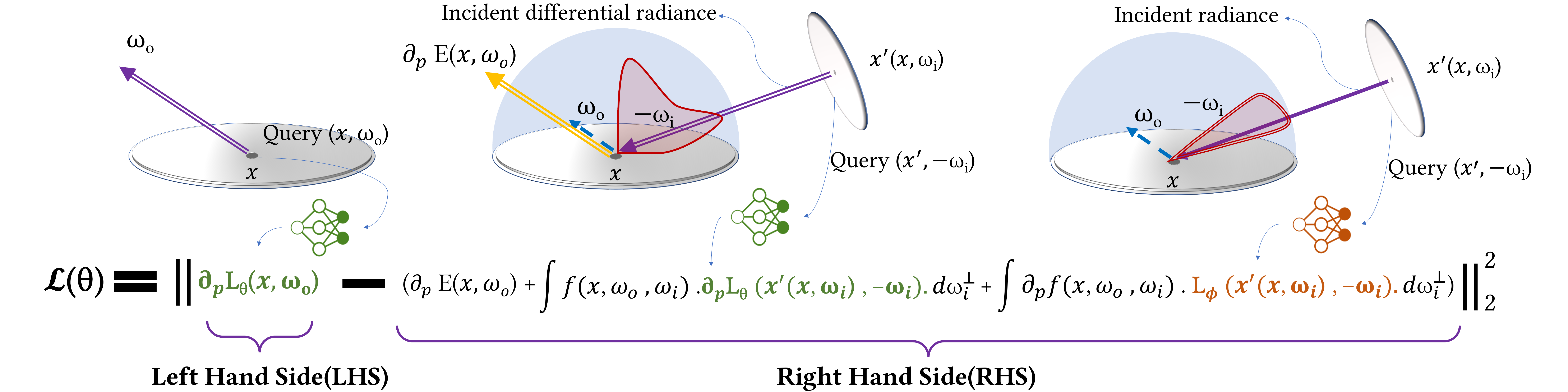}
    }
    \subcaptionbox{Our inverse optimization pipeline\label{fig:pipeline_inverse}}{  
    \includegraphics[width = \textwidth]{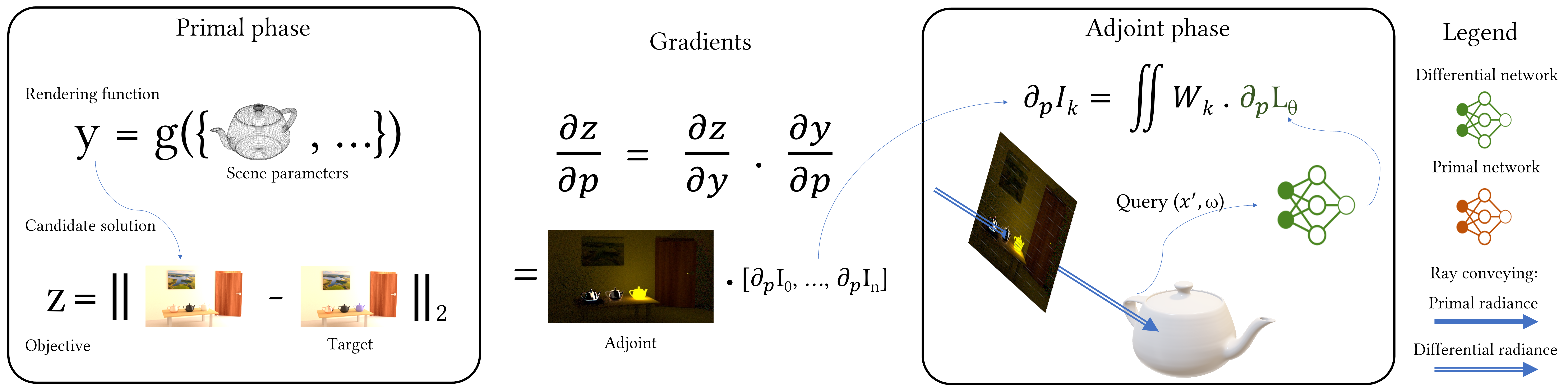}
    }
    \caption{Pipeline schematics illustrating our training scheme and inverse optimizations steps.}
\end{figure*}

Our training process has two steps: First, we train a neural network to learn the primal radiance distribution using Neural Radiosity \cite{hadadan2021neural}, i.e., minimizing the norm of the residual of the rendering equation in Equation \ref{eq:renderingresidual}. The resulting pre-trained $L_{\phi}$ then serves as an estimation of incident radiance in the residual of the differential rendering equation (Equations \ref{eq:differntial-rendering-equation} and \ref{eq:differntial-residual}). In order to train $\partial_p L_{\theta}$, the $L2$ norm of the differential residual is then minimized during training (see the loss function Equation~\ref{eq:differential-renderingloss}).

Similar to Neural Radiosity, our training scheme is a \emph{self-training} approach, that is, instead of providing noisy estimated data to our network as ground truth regression data, we compute both sides of the differential rendering equation using the same network and minimize the difference (residual) during training (Figure~\ref{fig:pipeline_training}).

\paragraph{Renderer} Our training scheme requires calls to rendering functions, automatic differentiable rendering functions, and deep learning optimization routines. For (differentiable) rendering purposes, we use Mitsuba2 \cite{Mitsuba2}, which provides useful Python bindings for the underlying C++ implementations, the ability to work with wavefronts of rays, and a differentiable megakernel that provides AD in backward and forward mode. We use Mitsuba AD to evaluate derivatives of $\partial_p E$ and $\partial_p f$ using forward mode. Mitsuba's megakernel allows parallel forward propagation for multiple parameters. 

\subsection{Sampling}

\paragraph{Norm of the Residual.}
To sample the norm of the residual, we need to sample locations $x_i$ and directions $w_{o,i}$ in Equation \ref{eq:lossMC}. Both variables are sampled on uniform distributions, i.e. $x$ uniformly on all the surfaces in the scene and $\omega_o$ uniformly on the unit hemisphere (in case of two-sided BSDFs, we sample on the unit sphere).

\paragraph{Hemispherical Integral.}
Monte Carlo estimation of the hemispherical scattering integrals in Equation \ref{eq:incidentintegral-MCestimation} requires sampling of incident directions $w_{i,j,k}$ and $w_{i,j,l}$. Firstly, we sample both integrals using the same set of incident directions to get performance gain. 
Since the two integrals are summed, the correlation between the two integrals using the same samples does not produce bias. Secondly, we importance sample emitter directions and BSDF distributions in \emph{primal space} to obtain incident directions. It is worth mentioning that our samples of $w_{i,j,k}$ are \emph{detached} \cite{MC_gradient_taxonomy}, meaning that they are static w.r.t perturbations in scene parameters. 

\subsection{Architecture}
Each of our networks is an MLP with 6 fully-connected layers with 512 neurons per hidden layer. The input parameters are the location $x$ and direction $\omega_o$. Inspired by Hadadan et al.~\shortcite{hadadan2021neural}, we also input extra surface properties such as the surface normal, and diffuse and specular reflectance at location $x$. Additionally, we also store multi-resolution learnable feature vectors in sparse grids in the scene and input them to the network, which empowers the network's ability to learn high frequency functions, demonstrated by Hadadan et al.~\shortcite{hadadan2021neural}. We adopt this approach for both of our networks. Recently, M\"uller et al.~\shortcite{muller2022instant} have introduced a multi-resolution hash encoding as an augmentation to neural networks that enables them to train in order of seconds, in addition to allowing the use of smaller networks without loss of quality. Adopting such architecture can enhance the efficiency of our method to use smaller networks with significant inference and training speedups.

\subsection{Updating Networks during Inverse Optimization}

As stated in Section \ref{inverse-redering}, the final step of inverse rendering involves fine-tuning after changes in the parameter space at every iteration. Ideally, each of our radiance and differential distributions should be updated to adapt to the scene parameters. 
That being said, we find fine-tuning a bottleneck during an optimization task. Instead, we propose to keep the networks intact in their initial state, but instead of fetching gradients from $\partial_p L_{\theta}$, we use the RHS of our solution (see the right hand side in Equation \ref{eq:differntial-rendering-equation}). The intuition behind this choice is that our RHS more quickly adapts to the scene parameter changes, as computing RHS requires one extra ray-tracing step using the updated parameters. After every update, terms $\partial f$ and $f$ in our scattering integrals used to evaluate the RHS are updated and current. While this leads to an additional bias in our gradients by keeping $\partial_p L_{\theta}$ and $ L_{\phi}$ outdated, our experiments show that we can still provide convergence improvements compared to the previous state-of-the-art.

\subsection{Differential Emission Re-parametrization}

Hadadan et al.~\shortcite{hadadan2021neural} show that taking the emission term in the residual  out of the network ($L_{\phi}$) makes the training process more stable. This is due to the high dynamic range of emitter values. Instead, they compute the emission term separately as it is a simple known function, and add it to the network output. We use a similar approach in the differential context, and the equations bellow reparametrize $\partial_{p}L_{\theta}$ w.r.t the differential emission term and substitute it in the operator form of the residual,
\begin{align}
\partial_{p}L_{\theta}(x,\omega_o) &= N_{\theta}(x,\omega_o) + \partial_{p}E(x,\omega_o),  \\
r_{\theta}(x,\omega_o) = \, &\partial_{p}L_{\theta}(x,\omega_o) - \partial_{p}E(x,\omega_o) -\mathrm{T}\{\partial_{p}L_{\theta}\}(x,\omega_o) \nonumber \\ &- T'\{L_{\phi}\}(x,\omega_o) \nonumber \\
= \, &N_{\theta}(x,\omega_o) -\mathrm{T}\{N_{\theta} + \partial_{p}E \}(x,\omega_o) \nonumber \\ &- T'\{N_{\phi} + E\}(x,\omega_o), \label{eq:emissionreparam}
\end{align}
where the transport operators are,
\begin{align}
\mathrm{T}\{\partial_{p}L_{\theta}\}(x,\omega_o) &= \int_{\mathcal{H}^2} f(x,\omega_i, \omega_o) \partial_{p}L_{\theta}(x'(x,\omega_i),-\omega_i) d\omega_i^{\perp} \\
\mathrm{T'}\{L_{\phi}\}(x,\omega_o) &= \int_{\mathcal{H}^2} \partial_{p}f(x,\omega_i, \omega_o) L_{\phi}(x'(x,\omega_i),-\omega_i) d\omega_i^{\perp}.
\end{align}

\newcommand\chairwidth{0.15}

\newcommand\mapeoffset{12pt}
\newcommand\mapeoffsett{26pt}

\begin{figure}
    \centering

    \hspace{5pt}
    \makebox[\chairwidth\textwidth]{{LHS}}
    \makebox[\chairwidth\textwidth]{{RHS}}
    \makebox[\chairwidth\textwidth]{{RB}}
    \vspace{5pt}

    
     \makebox[5pt]{\rotatebox{90}{\hspace{5pt} Copper roughness}}
     \includegraphics[trim={6cm 0 15cm 0},clip,width=\chairwidth\textwidth]{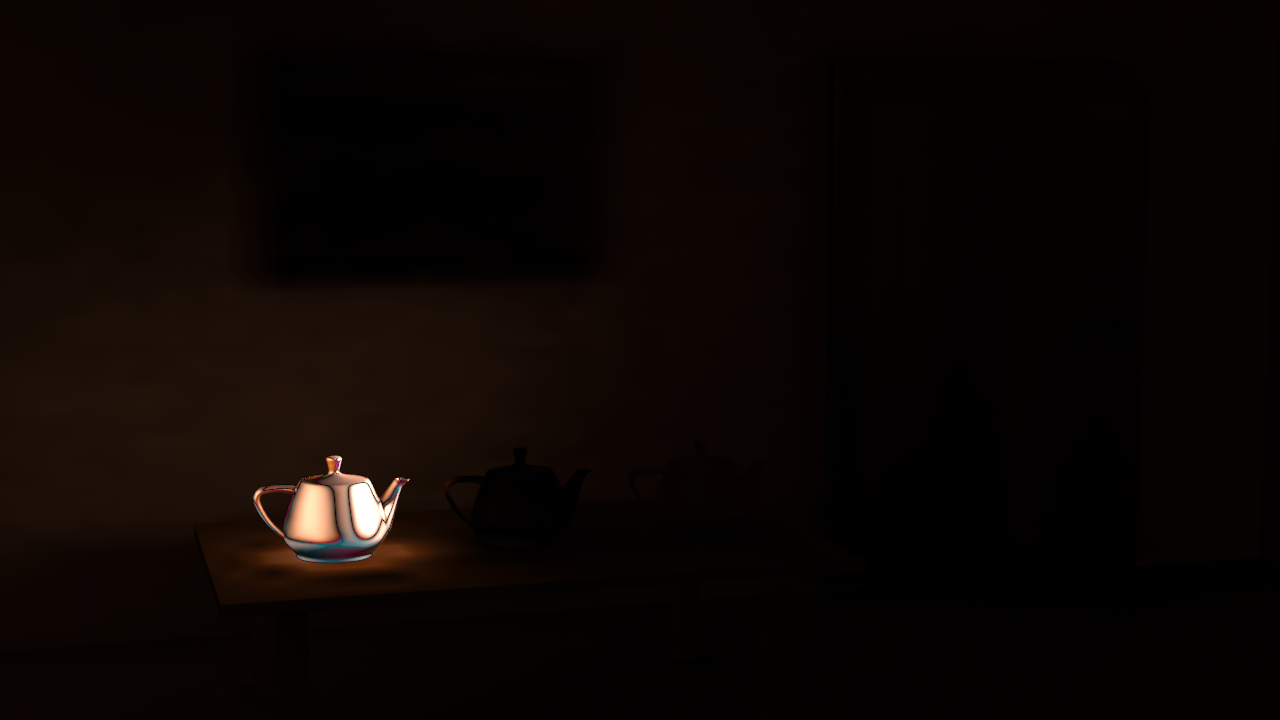}
     \includegraphics[trim={6cm 0 15cm 0},clip,width=\chairwidth\textwidth]{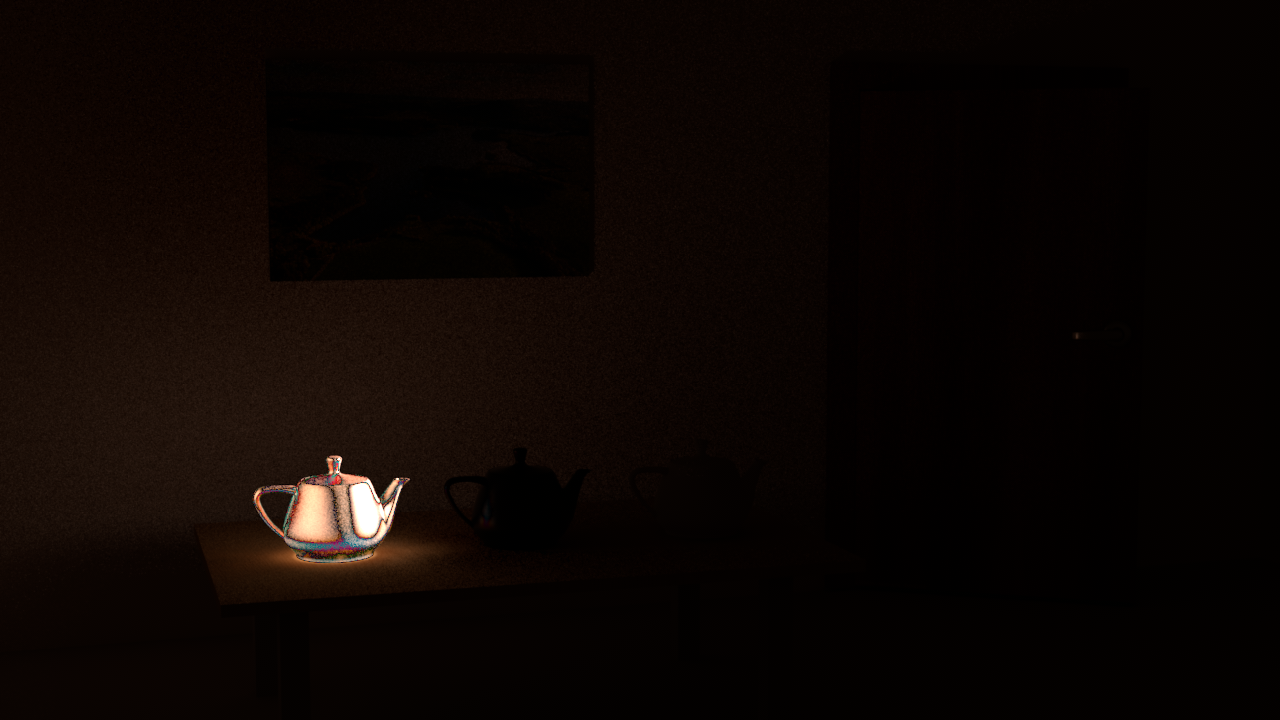}
     \includegraphics[trim={6cm 0 15cm 0},clip,width=\chairwidth\textwidth]{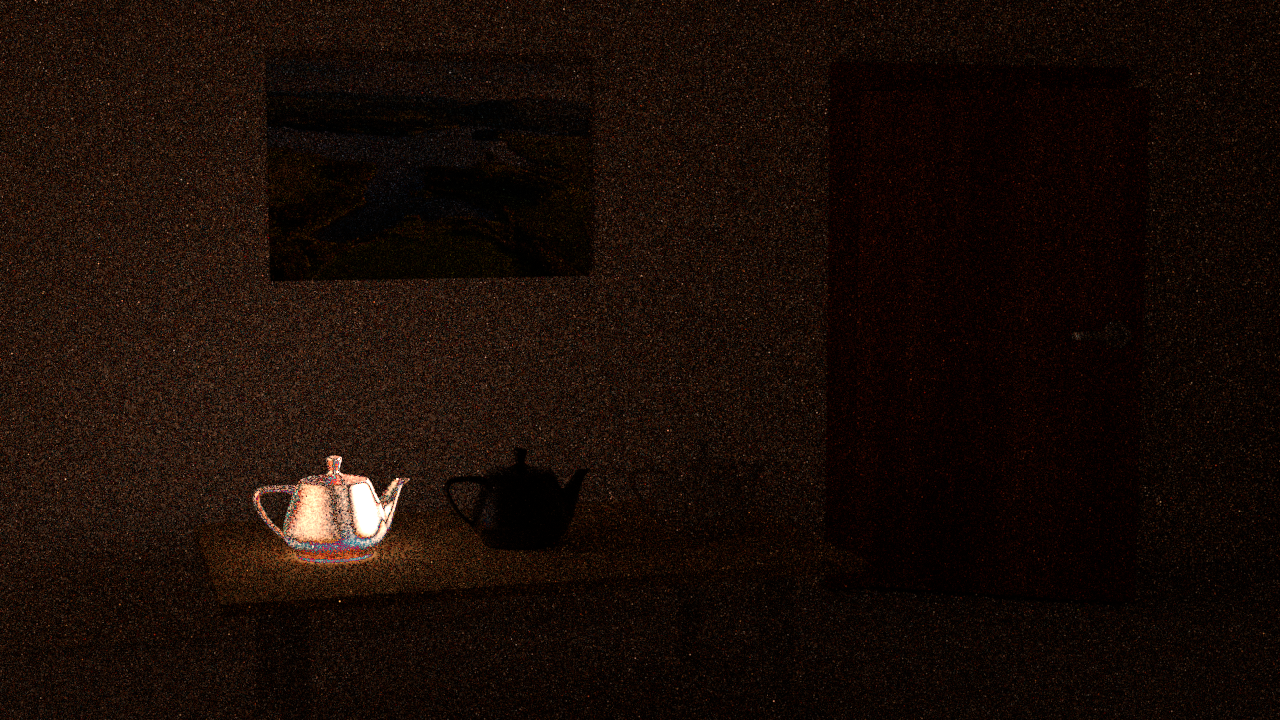}
     
     

     \makebox[5pt]{\rotatebox{90}{\hspace{3pt} Plastic roughness}}
     \includegraphics[trim={6cm 1cm 15cm 1cm},clip,width=\chairwidth\textwidth]{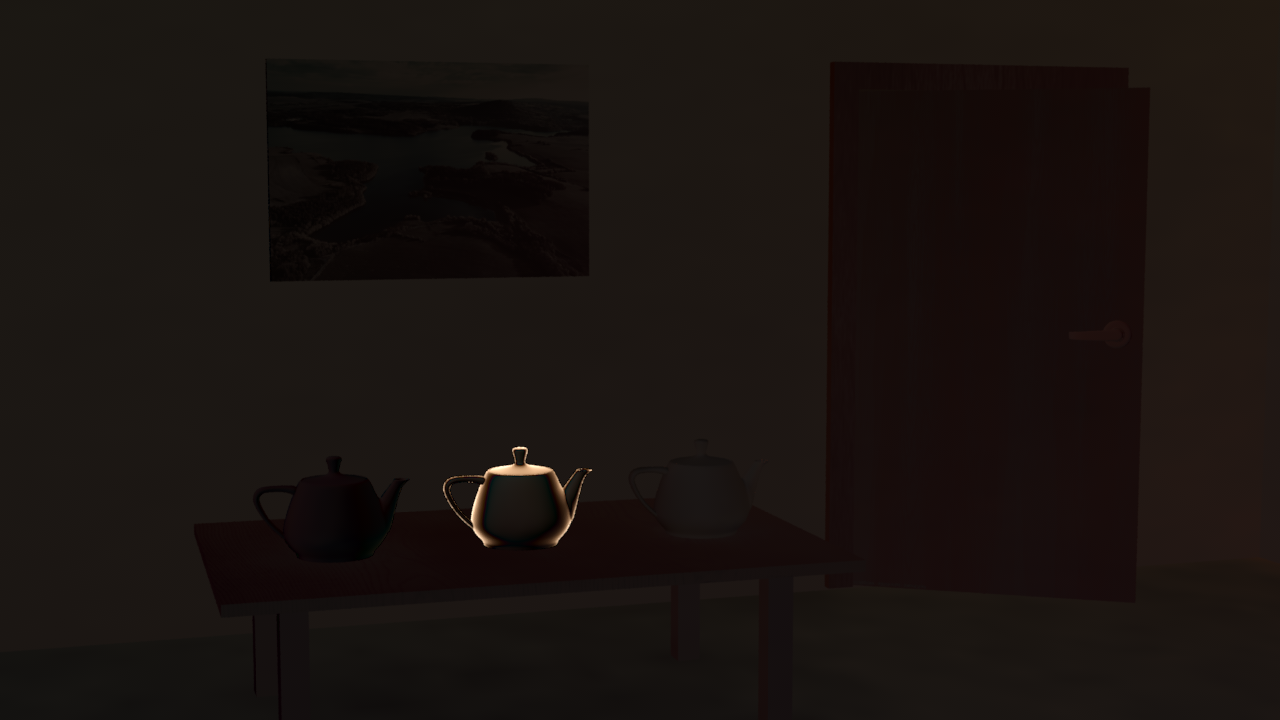}
     \includegraphics[trim={6cm 1cm 15cm 1cm},clip,width=\chairwidth\textwidth]{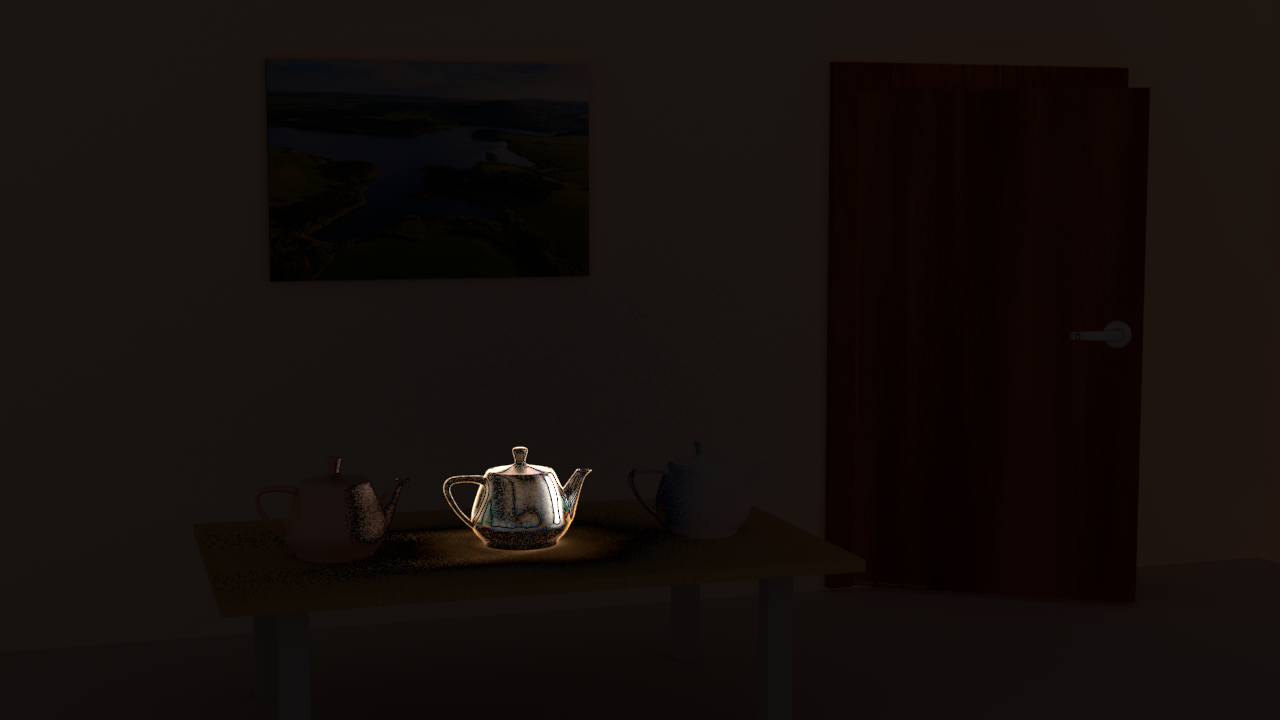}
     \includegraphics[trim={6cm 1cm 15cm 1cm},clip,width=\chairwidth\textwidth]{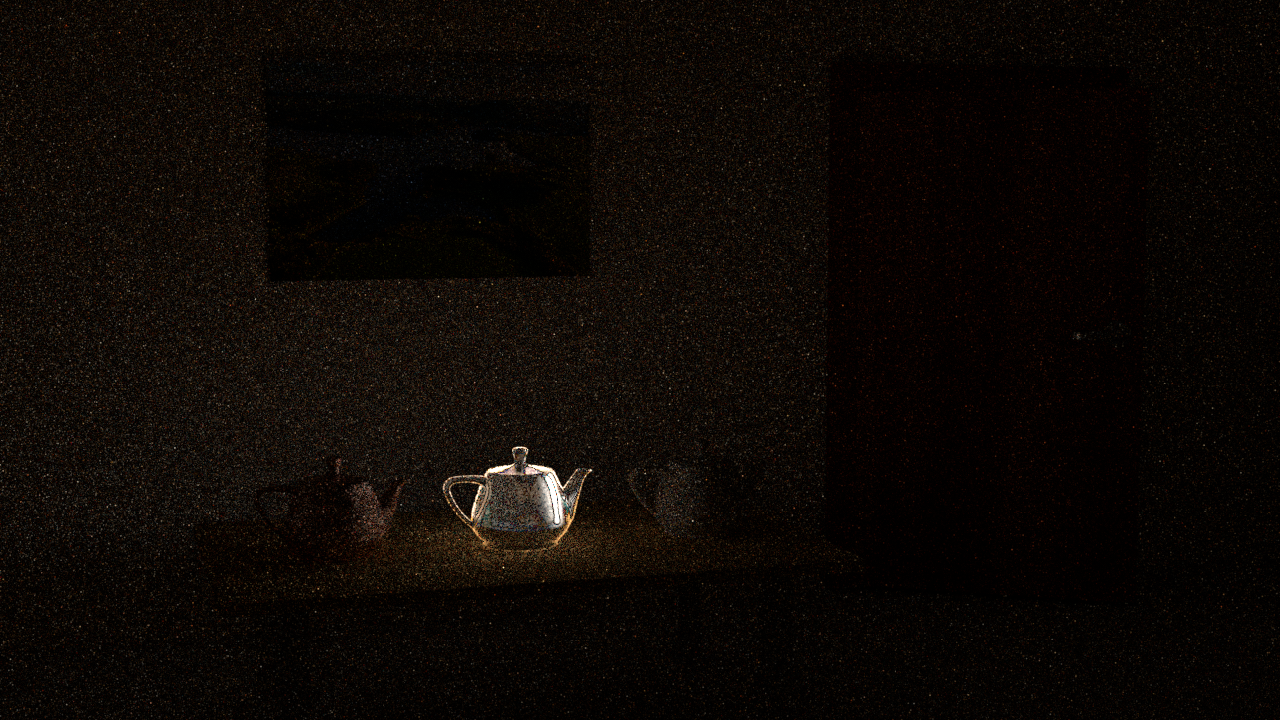}

     \makebox[5pt]{\rotatebox{90}{\hspace{10pt} Diffuse albedo}}
     \includegraphics[trim={6cm 1cm 15cm 1cm},clip,width=\chairwidth\textwidth]{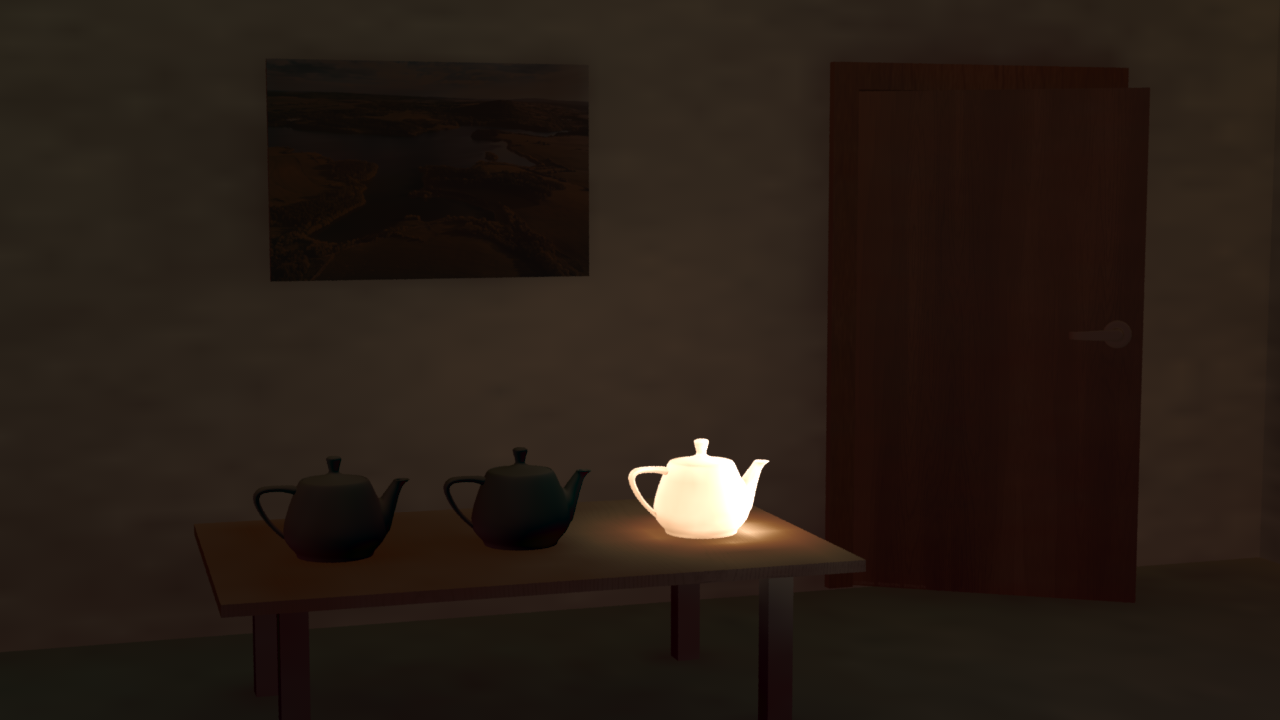}
     \includegraphics[trim={6cm 1cm 15cm 1cm},clip,width=\chairwidth\textwidth]{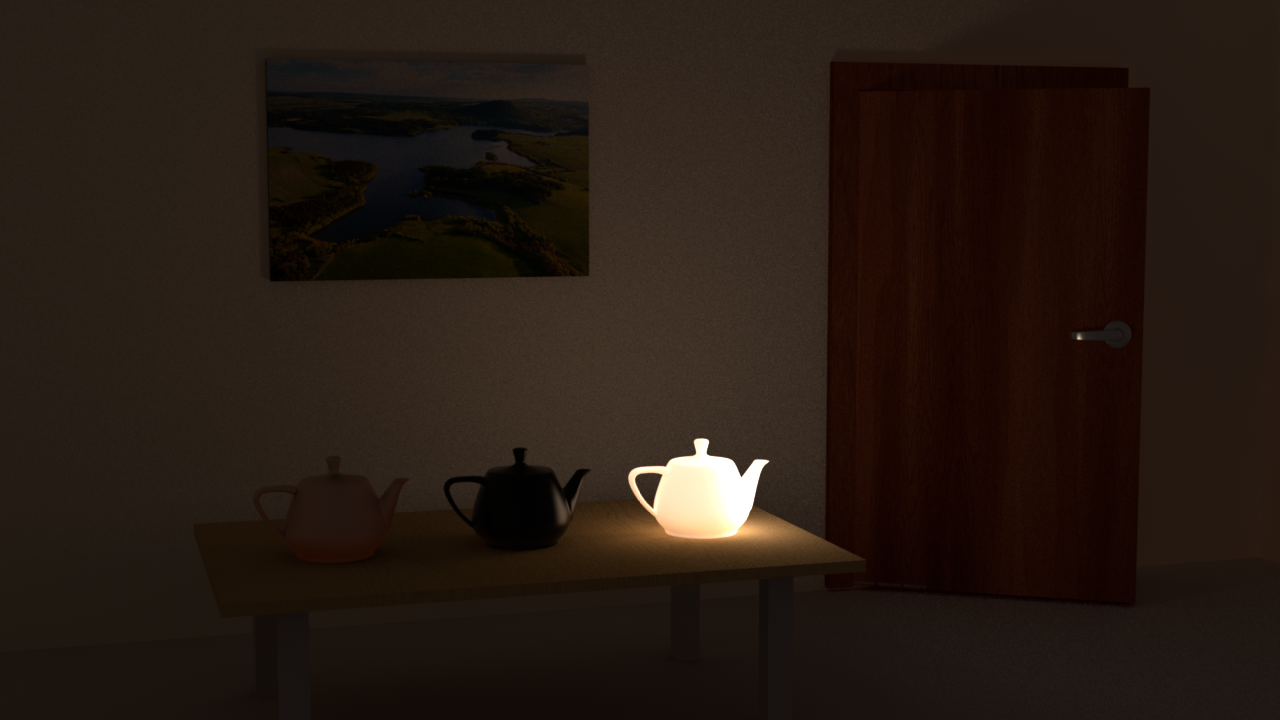}
     \includegraphics[trim={6cm 1cm 15cm 1cm},clip,width=\chairwidth\textwidth]{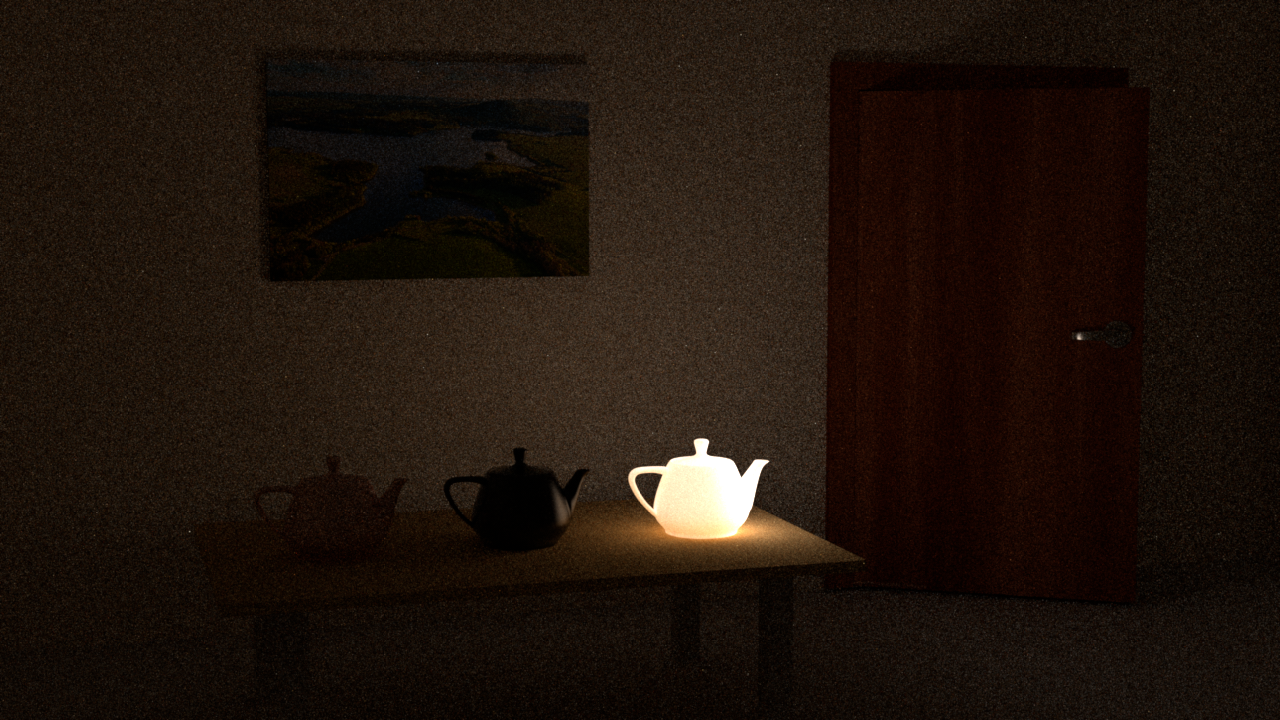}

    \caption{Separate renderings of our differential network w.r.t to the BRDF parameters of each teapot in the initial state in Figure~\ref{fig:teaser}. Our network is capable of learning non-diffuse gradients and global illumination effects in differential space. Using 8 spp for LHS and 2048 for RHS and RB.
    \label{fig:veach_grads}
    }
\end{figure}

\begin{figure*}[ht]
    \centering
    \includegraphics[width = \textwidth]{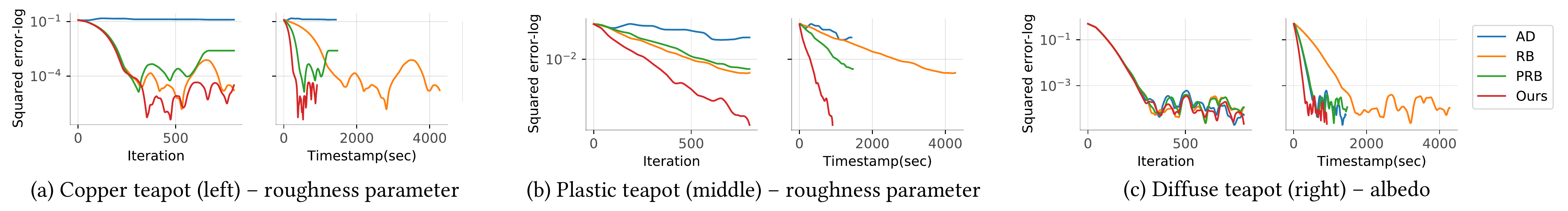}
    \caption{Experiment 1. Optimization curves for parameter error vs. iteration count and time (results in Figure \ref{fig:teaser}). Given a pre-trained network, our method achieves faster and better convergence compared to the baselines.}
    \label{fig:optim_curves}
\end{figure*}

\begin{figure*}[ht]
    \centering
    \includegraphics[width = \textwidth]{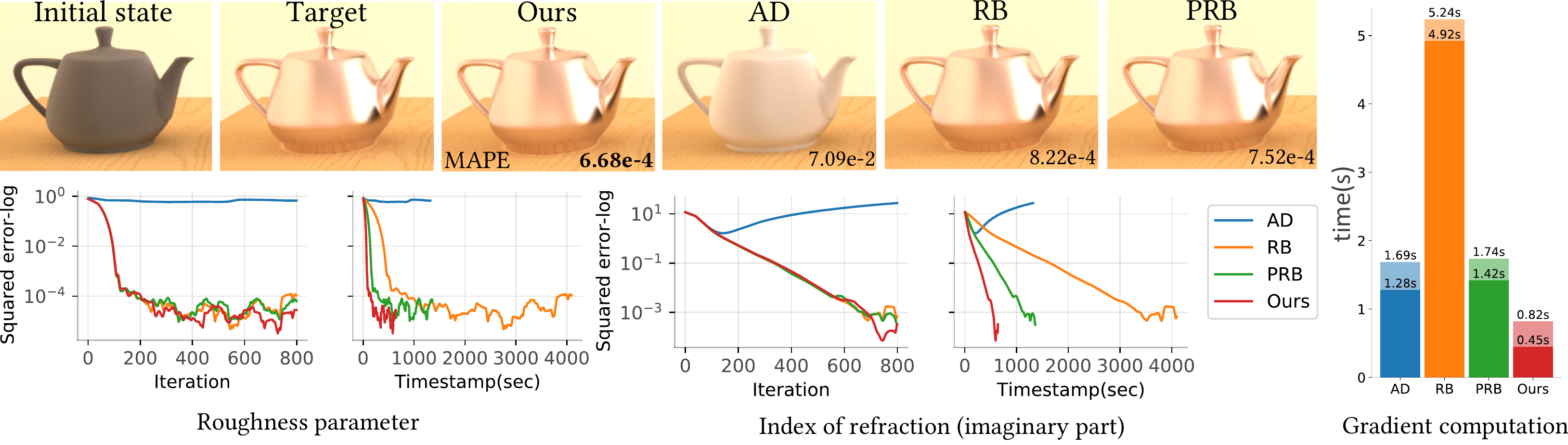}
    \caption{Experiment 2. Optimization for two BRDF parameters of one teapot. The initial state is designed such that the material behaves similar to a diffuse BRDF. Complete iteration time (higher number on each bar) and gradient computation time \emph{per iteration} (lower number) are reported. }
    \label{fig:optim_2}
\end{figure*}

\section{Results and analysis}

\subsection{Comparison to previous techniques}

In this section, we compare gradient computation for inverse rendering using our method to previous techniques, including generic AD, Radiative Backpropagation (RB), and its successor Path Replay Backpropagation (PRB):

\paragraph{AD} An inverse rendering optimization using AD requires computing $z(g(p))$ in a primal Monte Carlo rendering phase while recording a graph of all computations. 
Then, the evaluation of $\frac{\partial z}{\partial p}$ is simply propagating the derivatives directly from $z$ to the parameters. This process seems to be straightforward, but it has two main issues. First, as previously mentioned in \cite{AD_bias,radiative}, it produces biased gradients. This is because when estimating the expected value of Equation \ref{eq:inverse-chain-rule} using AD, both factors use the same set of Monte Carlo samples, and therefore $\mathbb{E}(ab) \neq  \mathbb{E}(a).\mathbb{E}(b)$ due to correlation between $a$ and $b$. 
The second issue is that storing the full computation graph of Monte Carlo rendering algorithms (and potentially a more descriptive objective) is memory intensive, making it prohibitive to use in complex scenes.
    
\paragraph{RB} To avoid recording the computation graph of generic Monte Carlo rendering, RB takes an adjoint approach to compute the gradients, vastly reducing the size of the  computation graph that needs to be stored. RB shares step (1) (in Section~\ref{inverse-redering}) with our pipeline. In step (2) and (3), RB finds $\frac{\partial y}{\partial p}$ (equivalently $\partial_p I_k$ and $\partial_p L$) using a special Monte Carlo path tracer that builds path integrals of differential radiance to solve the \emph{differential rendering equation}. Rays conveying an \emph{emitted adjoint} are traced from sensor pixels, and at every intersection, the accumulated differential light is absorbed by (backpropagated to) the scene parameters. Additionally, at every intersection, RB requires computing incident primal radiance by building another complete light path integral (observe that the second integral term in Equation~\ref{eq:differntial-rendering-equation} depends on incident $L$). This recursion leads to \emph{quadratic} complexity of RB in path length.

\paragraph{PRB} Path replay is similar to RB in spirit in the sense that it is adjoint-based and that it builds complete differential paths to solve the differential rendering equation. However, PRB finds the incident radiance more efficiently in a two-pass approach. In the primal pass, it first stores per-ray information that is used in a second adjoint pass that replays the first path using the same sequence of vertices and uses the information stored to precisely reconstruct the incident radiance at each step. Therefore, PRB solves the quadratic time complexity issue and is linear in path length.

\begin{table*}[t]
    \centering
    \caption{Details of inverse optimization experiments. All experiments use path tracing primal rendering at 16 spp and adjoint image sampling at 4 spp. We used different views of the sensor to render primal images in Experiments 1 and 2. A far view of Veach Door (in Figure~\ref{fig:teaser}a) was used at resolution 480x270 in Experiment 1. A close-up view (in Figure~\ref{fig:teaser}a) at resolution 480x120  was used for Experiment 2.}    
\begin{tabular*}{\textwidth}{c|c|c|c|c|c|c|c}
    Exp. & Fig. & \makecell[c]{Optimized \\ parameters} & \makecell[c]{Initial state} & \makecell[c]{Target state} & lr & \makecell[c]{Inverse optimization \\ peak memory-GiB} & \makecell[c]{Our model \\ training time} \\
    \hline
    1 & \ref{fig:teaser} \& \ref{fig:optim_curves} & \makecell[c]{Roughness \\ Roughness \\ Albedo}  & \makecell[c]{.4 \\ .1 \\ .8, .8, .8} & \makecell[c]{.05 \\ 0.4 \\ .3, .3, .9} & \makecell[c]{.0015 \\ .002 \\ .004}  & AD: 7.09, RB: 3.51, PRB: 3.00 Ours: 9.58 & \makecell[c]{$L_{\phi}$: 135m, $\partial_p L{\theta}$: 55m }\\
    \hline
    2 & \ref{fig:optim_2} & \makecell[c]{Roughness \\ Index of refraction} & \makecell[c]{1 \\ 1, 1, 1} & \makecell[c]{0.1 \\ 3.9, 2.4, 2.2} & \makecell[c]{.01 \\ .02}  & AD: 4.53, RB: 2.52, PRB: 2.46 Ours: 6.42  & \makecell[c]{$L_{\phi}$: 150m, $\partial_p L{\theta}$: 146m } \\
    \end{tabular*}

    \label{tab:optim_details}
\end{table*}

In comparison, our method provides the following benefits for inverse rendering described in Section~\ref{inverse-redering}:

\begin{itemize}[leftmargin=*]
    \item \textbf{Smooth gradients}: All three methods above compute gradients using Monte Carlo sampling, which produces noisy gradients. Instead, our networks $\partial_p L_{\theta}$ and $L_{\phi}$ produce smooth gradients that enables us to converge faster in optimization tasks.
    \item \textbf{Constant memory complexity}: AD's transcript requires storing per-bounce data to be able to invert the computation in the differentiable phase. For complex rendering tasks, the graph could easily grow larger than the available GPU memory. Instead, we take an adjoint method that consumes \emph{constant memory} in path length along with RB and PRB.
    \item \textbf{Constant time complexity}: 
    Given pre-trained networks, our time complexity is \emph{constant} in path length, since we do not build the complete path integral. Instead, we need to cheaply query our networks at the first intersection. Conversely, PRB and AD build complete light paths and are \emph{linear} in path length. Compared to RB, since we have incident primal radiance cached in $L_{\phi}$, our time complexity is not \emph{quadratic} in path length. 
    \item \textbf{View-independence}: Unlike all three methods above, our method provides view-independent solutions to the differential rendering equation (Figure~\ref{fig:multiview}). This means our solutions need not be recomputed/updated under changes of sensor parameters -- except if sensor parameters are in the set of parameters that are being optimized. This property can be a key improvement for multi-view optimization tasks.
 

    
\end{itemize}

\subsection{Inverse rendering optimization}
We conduct two inverse rendering optimizations using our method to find BRDF material properties in the \emph{Veach door} scene. This scene is of particular interest because the room is mostly lit by global illumination. Maximum path length is set to $15$ for the baselines. The three BRDF functions of the three teapots are rough conductor, rough plastic, and diffuse, from left to right. Please note that for optimization, we used path tracing for primal rendering across all baselines as well as in our method. Our models were only trained for the initial states in each experiment.

\paragraph{Experiment 1.}

In our first experiment, we optimize one BRDF parameter per teapot. Table \ref{tab:optim_details} lists the optimized parameters in this experiment, and their initial and target states. The solution of our differential network for the initial state can be found in Figure~\ref{fig:veach_grads}. From Figure~\ref{fig:teaser}, we can observe that our optimization method is capable of finding near specular target states (for the left teapot), as well as the opposite direction for the middle teapot. Also, it shows that our method improves both gradient computation speed and accuracy of the final results compared to the baselines. Figure \ref{fig:optim_curves} plots the optimization curves for this optimization, indicating that we achieve faster and better convergence than the other methods.

\paragraph{Experiment 2.}

In the second experiment, we optimize two BRDF parameters for the left teapot, its roughness and the imaginary part of the index of refraction. The initial state in this experiment is designed such that the BRDF initial state appears similar to a diffuse material. Figure \ref{fig:optim_2} shows that we can successfully achieve more accurate final results compared to the reference target state, in addition to faster convergence and speed up in gradient computation time in comparison with baselines. We refer to Table \ref{tab:optim_details} for details of this experiment.

\section{Conclusions and Future Work}

In this paper, we introduced a new method to solve the differential rendering equation using a single neural network. Our network parameters are optimized directly by minimizing the norm of the residual of the differential rendering equation. Our learnable network architecture is capable of representing the full continuous, view-independent differential radiance distribution and accounts for global differential illumination. A pre-trained instance of our network can be leveraged in inverse optimization problems to reconstruct complex non-diffuse BRDF properties and achieve faster convergence with better time and memory complexity than the state of the art. Finally, the view-independence of our solution has the potential to produce further speedups in multi-view optimizations.

Our method has a number of limitations. First, our differential network requires an output channel for each scene parameter included in the gradient. As we use fully-connected layers, the number of connections in the last layer grows exponentially with the number of outputs and this generates a memory constraint for differentiable rendering tasks that require optimizing w.r.t millions of parameters. Additionally, our training/fine-tuning process currently is time consuming, but could be improved in many ways. For example, our residual sampling uses a uniform distribution. A more efficient implementation could use importance sampling instead. Further, we use importance sampling of emitters and BSDFs in the primal space. A more efficient method would be to importance sample incident directions according to the underling \emph{differential} emission ($\partial_p E$) and BSDF ($\partial_p f$) distribution. Such a modification would produce less noise when evaluating the RHS, which would result in smoother signals during both training and inverse optimization. Finally, we could use high-performance implementations of fully connected networks instead of standard Pytorch functionality, which can provide large speedups as shown in previous work.

\bibliographystyle{ACM-Reference-Format}
\bibliography{references}


\end{document}